\documentclass[prl,twocolumn,nofootinbib,superscriptaddress]{revtex4-1}

\usepackage{graphicx}
\usepackage{dcolumn}
\usepackage{bm}
\usepackage{slashed}
\usepackage{amsmath,graphicx}
\usepackage[colorlinks=true,linktocpage=true,linkcolor=blue,citecolor=blue]{hyperref}
\usepackage{float}
\usepackage{nicefrac}
\usepackage[normalem]{ulem}
\usepackage{amsmath}
\usepackage{subfigure}

\newcommand{\beq}{\begin{eqnarray}}
\newcommand{\eeq}{\end{eqnarray}}

\newcommand{\bel}[1]{\begin{eqnarray}\label{#1}}
\newcommand{\eel}{\end{eqnarray}}

\newcommand{\rf}[1]{Eq.~(\ref{#1})}
\newcommand{\rfm}[1]{Eqs.~(\ref{#1})}
\newcommand{\rftwo}[2]{Eqs.~(\ref{#1})~and~(\ref{#2})}
\newcommand{\rfn}[1]{~(\ref{#1})}

\newcommand{\nn}{\nonumber}

\newcommand{\p}{\partial}

\newcommand{\tr}{\rm tr}

\newcommand{\f}[2]{\frac{#1}{#2}}
\newcommand{\onehalf}{{\nicefrac{1}{2}}}
\newcommand{\half}{\f{1}{2}}

\renewcommand\sout{\bgroup \color{blue} \ULdepth=-.5ex \ULset}
\newcommand{\ed}{{\varepsilon}}       % energy density
\def\gmunu{g^{\mu\nu}}

\def\TmnU{T^{\mu \nu}}

\def\n0{n_{(0)}}
\def\e0{\varepsilon_{(0)}}
\def\P0{P_{(0)}}
\def\s0{s_{(0)}}

\def\fplusrsxp{f^+_{rs}(x,p)}

\def\bmu{\beta_\mu}

\def\umU{u^\mu}  
                         
\def\umL{u_\mu}

\def\unu{u^\nu}

\def\unuL{u_\nu}

\def\pmu{p^\mu}
\def\pnu{p^\nu}

\def\omnL{\omega_{\mu\nu}}
\def\omnU{\omega^{\mu\nu}}
\def\omnLbar{{\bar \omega}_{\mu\nu}}
\def\omnUbar{{\bar \omega}^{\mu\nu}}

\def\oabU{\omega^{\alpha\beta}}
\def\omnLD{{\tilde \omega}_{\mu\nu}}

\def\epsLmnbg{\epsilon_{\mu\nu\beta\gamma}}

\def\epsLmnab{\epsilon_{\mu\nu\alpha\beta}}

\def\kmL{k_\mu}
\def\kbar{\bar k}
\def\kmLbar{{\bar k}_\mu}

\def\knL{k_\nu}

\def\omg{\omega^\gamma}
\def\obar{\bar\omega}

\def\omL{\omega_\mu}

\def\omLbar{{\bar \omega}_\mu}

\def\SmunuU{{\Sigma}^{\mu\nu}}

\def\ubarrp{{\bar u}_r(p)}

\def\usp{u_s(p)}
\def\urp{u_r(p)}

\def\vbarrp{{\bar v}_r(p)}
\def\vbarsp{{\bar v}_s(p)}
\def\vsp{v_s(p)}
\def\vrp{v_r(p)}

\def\g5{\gamma_5}

\def\fminusrsxp{f^-_{rs}(x,p)}

\def\slmnU{S^{\lambda, \mu \nu}}

\def\Ot{\tilde \Omega}

\def\mmL{m_\mu}
\def\mmU{m^\mu}
\def\nmL{n_\mu}
\def\nmU{n^\mu}

\def\mnL{m_\nu}

\def\nnL{n_\nu}

\begin{document}
%%%%%%%%%%%%%%%%%%%%%%%%%%%%%%%%%%%%%%%%%%%%%%
%%%%%%%%%%%%%%%%%%%%%%%%%%%%%%%%%%%%%%%%%%%%%%

%%%%%%%%%%%%%%%%%%%%%%%%%%%%%%%%%%%%%%%%%%%%%%%%%%%%%% 
 
\title{Relativistic fluid dynamics with spin}

\author{Wojciech Florkowski} 
\affiliation{Institute of Nuclear Physics Polish Academy of Sciences, PL-31342 Krakow, Poland}
\affiliation{Jan Kochanowski University, PL-25406 Kielce, Poland}
\affiliation{ExtreMe Matter Institute EMMI, GSI, D-64291 Darmstadt, Germany}
\author{Bengt Friman} 
\affiliation{GSI Helmholtzzentrum f\"ur Schwerionenforschung, D-64291 Darmstadt, Germany}
\author{Amaresh Jaiswal} 
\affiliation{GSI Helmholtzzentrum f\"ur Schwerionenforschung, D-64291 Darmstadt, Germany}
\affiliation{School of Physical Sciences, National Institute of Science Education and Research, HBNI, Jatni-752050, India}
\author{Enrico Speranza} 
\affiliation{GSI Helmholtzzentrum f\"ur Schwerionenforschung, D-64291 Darmstadt, Germany}
\affiliation{Institut f{\"ur} Kernphysik, Technische Universit{\"a}t Darmstadt, D-64289 Darmstadt, Germany}
\date{\today}

\begin{abstract}

Using the conservation laws for charge, 
energy, momentum, and angular momentum, we derive hydrodynamic equations 
for the charge density,  local temperature, and fluid velocity, as well as for 
the polarization tensor, starting from local equilibrium distribution functions for particles and antiparticles
with spin~$\onehalf$. The resulting set of 
differential equations extend the 
standard picture of perfect-fluid hydrodynamics with a conserved entropy 
current in a minimal way. This framework can be used in space-time analyses of the evolution of spin 
and polarization in various physical systems including 
high-energy nuclear collisions. We demonstrate that a stationary 
vortex, which exhibits vorticity-spin alignment, corresponds 
to a special solution of the spin-hydrodynamical equations. 
 
\end{abstract}



\maketitle 
%
%%%%%%%%%%%%%%%%%%%%%%%%%%%%%%%%%%%%%%%%%%%%%%%%%%%%%%%%%%%%%%%%%%%%%%%%
%\section{Introduction}
%\label{sect:intro}
%%%%%%%%%%%%%%%%%%%%%%%%%%%%%%%%%%%%%%%%%%%%%%%%%%%%%%%%%%%%%%%%%%%%%%%%

{\it 1. Introduction:} Non-central heavy-ion collisions create 
fireballs with large global angular momenta which may generate a spin
polarization of the hot and dense matter in a way resembling the equilibrium magnetomechanical effects of 
Einstein and de Haas~\cite{dehaas:1915} and Barnett~\cite{Barnett:1935}. 
Consequently, much effort has recently been invested in studies of polarization and spin 
dynamics of particles produced in high-energy nuclear collisions, 
both from the experimental and theoretical point of view (for a recent review
see~\cite{Wang:2017jpl}). In this context, theoretical studies have explored
the role of the spin-orbit coupling~\cite{Betz:2007kg,Liang:2004ph,Liang:2004xn,Gao:2007bc,Chen:2008wh} and 
global equilibrium of rotating bodies~\cite{Becattini:2009wh,Becattini:2012tc,
Becattini:2013fla,Becattini:2015nva,Hayata:2015lga}. See also the kinetic 
models of spin dynamics~\cite{Gao:2012ix,Chen:2012ca,Fang:2016vpj,Fang:2016uds,Fang:2016vpj}
and the works on anomalous hydrodynamics~\cite{Son:2009tf,Kharzeev:2010gr}. 
Recent work, based on the Lagrangian formulation of hydrodynamics, is reported in Refs.~\cite{Montenegro:2017rbu,Montenegro:2017lvf}.
Moreover, it has been suggested that the global angular momentum should be reflected in the polarization of observed hadrons, e.g., in the case
of $\Lambda$ hyperons and vector mesons \cite{Liang:2004ph,Liang:2004xn,
Becattini:2016gvu, STAR:2017ckg}.

Surprisingly, no dynamical fluid-like framework has been developed so far 
which allows for space-time evolution of polarization effects, despite the 
fact that the studies of fluids with spin have a long history 
initiated already in 1930's \cite{Mathisson,Weyssenhoff}. Recent 
works have contributed to our understanding of global 
equilibrium/stationary states, which exhibit interesting features of 
vorticity-spin alignment \cite{Becattini:2009wh,Becattini:2015nva}. 
Moreover, polarization effects present during the final freeze-out 
stage of collisions, where particles cease to interact, have also been 
studied \cite{Becattini:2013vja, Becattini:2013fla,Fang:2016vpj}. On 
the other hand, little is known about the changes of polarization 
during the collision process, especially, if the latter is described 
with the help of fluid dynamics.

In this work, we develop a general perfect-fluid framework for 
charged particles with spin $\onehalf$, which allows for space and 
time dependent studies of polarized fluids. We show how the 
conservation laws for the charge current, energy-momentum, and 
angular momentum can be used consistently to obtain the
dynamics of fluids with spin degrees of freedom. In addition to the standard 
equations, which determine the time dependence of the 
charge density $n$, temperature $T$  and fluid four-velocity $\umU$, 
our equations determine the dynamics of the polarization tensor $\omnU$. The 
evolution equation for polarization presented here provides a consistent 
theoretical framework for studying the effects of vorticity and polarization 
in hydrodynamic simulations of heavy-ion collisions. Furthermore, our approach 
is applicable also to other systems exhibiting collision dominated, 
collective behavior connected with non-trivial polarization effects.

We note at this point that in ideal hydrodynamics, the system is 
assumed to be in local thermodynamic equilibrium. Thus, the 
polarization of each fluid cell equals the local equilibrium value, specified 
by the thermodynamic variables. 
Consequently, the ideal fluid-dynamic framework presented here, 
does not account for the relaxation of spin degrees of freedom, 
e.g., through the spin-orbit interaction~\cite{HessWaldmann}.

Below we use the following conventions and notation for the metric tensor, 
Levi-Civita's tensor, and the scalar product: $g_{\mu\nu} = 
\hbox{diag}(+1,-1,-1,-1)$, $\epsilon^{0123} = -\epsilon_{0123} =~1$, 
$a \cdot b = g_{\mu \nu} a^\mu b^\nu$. Throughout the text we use
natural units with $c = \hbar = k_B =1$.

%\medskip
{\it 2. Local distribution functions:} Our starting point is the 
phase-space distribution functions for spin-$\onehalf$ particles and 
antiparticles in local equilibrium, as introduced in \cite
{Becattini:2013fla}. In order to incorporate the spin degrees of 
freedom, they are generalized from scalar functions to two by 
two spin density matrices for each value of the space-time position $x$ 
and momentum $p$,
\bel{fplusrsxp}
\fplusrsxp = \f{1}{2m} \ubarrp X^+ \usp,
\eel
\bel{fminusrsxp}
\fminusrsxp = - \f{1}{2m} \vbarsp X^- \vrp.
\eel
Here $m$ is the particle mass and $\urp$ and $\vrp$ are bispinors (with the spin indices $r$ and 
$s$ running from 1 to 2), with the normalization $\ubarrp \usp=2m\,\delta_{rs}$ and $\vbarrp \vsp=-2 m\,\delta_{rs}$. Note the minus 
sign and different ordering of spin indices in\rfn{fminusrsxp}
compared to\rfn{fplusrsxp}.

Following the notation used in \cite{Becattini:2013fla}, we introduce the 
matrices
\bel{XpmM}
X^{\pm} =  \exp\left[\pm \xi(x) - \bmu(x) \pmu \right] M^\pm, 
\eel
where
\bel{Mpm}
M^\pm = \exp\left[ \pm \f{1}{2} \omnL(x)  \SmunuU \right] .
\eel
In these equations, we use the notation $\beta^\mu= \umU/T$ and $\xi = \mu/T$, 
with the temperature $T$, chemical potential $\mu$ and four velocity $\umU$. 
The latter is normalized to $u^2~=~1$. Moreover, $\omnL$ is the polarization tensor, while  $\SmunuU$ 
is the spin operator expressed in terms of the Dirac gamma matrices, $\SmunuU 
= (i/4) [\gamma^\mu,\gamma^\nu]$.  
For the sake of simplicity, we restrict ourselves to classical Boltzmann statistics in this 
work. However, given the closed form expression for $M^\pm$ obtained 
below, it is a straighforward exercise to generalize our discussion to Fermi-Dirac statistics.

The antisymmetric polarization tensor $\omnL$ can be represented by the following tensor decomposition
\bel{omunuL}
\omnL \equiv \kmL \unuL - \knL \umL + \epsLmnbg u^\beta \omg .
\eel
We note that any part of $k_\mu$ or $\omega_\mu$ that is parallel to $\umU$, is cancelled in\rfn{omunuL}. Hence, we can assume that both $k_\mu$ and 
$\omega_\mu$ are orthogonal to $\umU$, i.e., $k \cdot u = \omega \cdot  u = 0$, and express the four-vectors $k_\mu$ 
and $\omega_\mu$ in terms of $\omnL$ using
\bel{kmuomu}
k_\mu = \omnL \unu, \quad \omega_\mu = \f{1}{2} \epsLmnab \, \omega^{\nu\alpha} u^\beta.
\eel
This means that  $k_\mu$ and $\omega_\mu$ are space-like 
four-vectors with only three independent components.  

It is convenient to introduce the dual polarization tensor
\bel{omunuLD}
\omnLD \equiv \f{1}{2} \epsLmnab  \oabU = \omega_\mu \unuL - \omega_\nu \umL +  \epsLmnab k^\alpha u^\beta.
\eel
Using \rftwo{omunuL}{omunuLD} one finds that the scalar contraction of the polarization 
tensor with itself gives $\f{1}{2} \omnL \omnU = k \cdot k - \omega 
\cdot \omega$, whereas the contraction of the polarization  tensor with its 
dual yields $\f{1}{2} \omnLD \omnU = 2 k \cdot \omega$. 

The exponential function in \rf{Mpm} is defined in terms of a power series, which can be resummed
(most easily in the chiral representation of the $\gamma$ matrices,
where $\SmunuU$ is block diagonal). Using the constraint
\bel{conONE}
k \cdot \omega  = 0
\eel
we find the compact form
\bel{Mpmexp}
M^\pm &=& \cosh(\zeta) \pm  \f{\sinh(\zeta)}{2\zeta}  \, \omnL \SmunuU  ,
\eel
where
\bel{zeta}
\zeta \equiv \f{1}{2} \sqrt{ k \cdot k - \omega \cdot \omega }.
\eel
We now assume that $k \cdot k - \omega \cdot \omega \geq 0$, which in conjunction with \rf{zeta} implies that $\zeta$ is real. We motivate these choices below, following Eqs.\rfn{prs} and\rfn{dP}.

%\medskip
{\it 3. Basic observables:}
Using the distribution functions (\ref{fplusrsxp}) and (\ref{fminusrsxp}), we obtain the basic 
hydrodynamic quantities. The charge current is given by~\cite{deGroot:1980}
\bel{jmu}
\hspace{-0.5cm} N^\mu &=&  \int \f{d^3p}{2 (2\pi)^3 E_p}  \pmu \left[ \tr( X^+ ) - \tr ( X^- )  \right] = n \umU,
\eel
where ``$\tr$'' denotes the trace over spinor indices and
\bel{nden}
n = 4 \, \cosh(\zeta) \sinh(\xi)\, \n0
\eel
is the charge density. Here $\n0(T) = \langle(u\cdot p)\rangle_0$ is 
the number density of spin 0, neutral Boltzmann particles, obtained using the thermal average
\bel{avdef}
\langle \cdots \rangle_0 \equiv \int \f{d^3p}{(2\pi)^3 E_p}  (\cdots) \,  e^{- \beta \cdot p} ,
\eel
where $E_p = \sqrt{m^2 + {\boldsymbol p}^2}$.

The energy-momentum tensor for a perfect fluid then has the form~\cite{deGroot:1980}
\bel{Tmn}
\TmnU &=&  \int \f{d^3p}{2 (2\pi)^3 E_p}  \pmu \pnu \left[ \tr( X^+ ) +  \tr ( X^- )  \right] \nn \\
&=& (\varepsilon + P ) \umU \unu - P \gmunu,
\eel
where the energy density and pressure are given by
\bel{enden}
\varepsilon = 4 \, \cosh(\zeta) \cosh(\xi) \, \e0 
\eel
and
\bel{prs}
P = 4 \, \cosh(\zeta) \cosh(\xi) \, \P0,
\eel
respectively. In analogy to the density $\n0(T)$, we define the auxiliary 
quantities $\e0(T) = \langle(u\cdot p)^2\rangle_0$ and $\P0(T) = -(1/3) \langle \left[ p\cdot p - 
(u\cdot p)^2 \right] \rangle_0$.  At this point we note that in the case where 
$\zeta$ is not real, one can find a generalized form of $M^\pm$ and consequently of all thermodynamic
quantities, involving trigonometric functions. As this potentially leads to 
negative values of the pressure, we exclude such cases
from the present investigation.  We also note that  the energy-momentum tensor (\ref{Tmn}) is symmetric, 
owing to the fact that we deal with classical particles that have a well defined relation
between energy, momentum and velocity, ${\boldsymbol p} = E_p \,{\boldsymbol v}$~\cite{Hehl:1976vr}.

The entropy current is given by an obvious generalization of the Boltzmann 
expression
\bel{s2}
S^\mu &=&  -\int \f{d^3p}{2 (2\pi)^3 E_p}  \, \pmu  \, \Big( \tr\left[ X^+ (\ln X^+ -1)\right] \nn \\
&& \hspace{2cm}  +  \, \tr \left[ X^- (\ln X^- -1) \right] \Big).
\eel
This leads to the following  entropy density
\bel{s}
s = u_\mu S^\mu = \f{\ed+P  - \mu \, n - \Omega w}{T} ,
\eel
where $\Omega$ is defined through the relation $\zeta = \Omega/T$ and 
\bel{w}
w = 4 \, \sinh(\zeta) \cosh(\xi) \, \n0.
\eel
Equation\rfn{s} suggests that $\Omega$ should be used as a thermodynamic variable of the 
grand canonical potential, in addition to $T$ 
and $\mu$. Taking the 
pressure $P$ to be a function of $T, \mu$ and $\Omega$, we find
\bel{dP}
s = \left.{\f{\p P}{\p T}}\right\vert_{\mu,\Omega}, \quad 
n = \left.{\f{\p P}{\p \mu}}\right\vert_{T,\Omega}, \quad 
w = \left.{\f{\p P}{\p \Omega}}\right\vert_{T,\mu}.
\eel
The modified thermodynamic relation\rfn{s} is analogous to the one obtained in 
Ref.~\cite{Becattini:2009wh}. We note  that the thermodynamic 
relations\rfn{dP} also suggest that $\zeta$ should be real. 

Moreover, we observe that the thermodynamic variable $\Omega$ 
controls the polarization of the system. Hence, in the present framework, 
$\Omega$ acts as a proxy for the spin-vorticity coupling, which provides 
a spin-dependent shift in the single-particle energies~\cite{Hehl:1990nf}.   
In global equilibrium, $\Omega$ 
is a unique function of the thermal vorticity~\cite{Becattini:2013fla}. 
However, in local equilibrium and, in particular, in non-equilibrium 
systems, this relation may be relaxed. 

In this paper we explore the dynamics of systems, where the
local polarization and thermal vorticity are initialized as independent variables and evolve according to ideal hydrodynamics. Thus, we allow for an incomplete relaxation of the spin degrees of freedom during the pre-hydrodynamic stage, while in the hydrodynamic phase dissipative processes, in particular also spin relaxation, are neglected. We stress that this idealized framework allows for non-trivial spin dynamics, and hence provides the possibility to perform key studies of polarization phenomena in a hydrodynamic setting. 

%\medskip
{\it 4. Basic conservation laws:} 
Before we turn to the discussion of the 
spin observables let us analyze the basic conservation laws. The 
conservation of energy and momentum requires that
\bel{Tmncon1}
\p_\mu \TmnU = 0.
\eel
This equation can be split into  two parts, one longitudinal and the other transverse with
respect to $u^\mu$:
\bel{Tmncon2}
\p_\mu [(\ed + P) \umU ] &=& \umU \p_\mu P \equiv \f{dP}{d\tau}, \nn \\
 (\ed + P ) \f{d \umU}{d\tau} &=& (g^{\mu \alpha} - u^\mu u^\alpha ) \p_\alpha P.
\eel
Evaluating the derivative on the left-hand side of the first equation in\rfn{Tmncon2} 
and using\rfn{dP} we find
\bel{snwcon}
T \,\p_\mu (s \umU) + \mu \,\p_\mu (n \umU) + \Omega \,\p_\mu (w \umU) = 0.
\eel
The middle term in \rf{snwcon} vanishes due to charge
conservation,
\bel{ncon}
\p_\mu (n \umU)=0.
\eel
Thus, in order to have entropy conserved in our system (for the 
perfect-fluid description we are aiming at), we demand that
\bel{wcon}
\p_\mu (w \umU) = 0.
\eel
Consequently, using \rf{ncon} and \rf{wcon} we self-consistently 
arrive at the equation for conservation of entropy,  $\p_\mu (s \umU)~=~0$.

Note that in the absence of a net spin polarization, i.e., for $\zeta~=~0$,  
\rf{nden} reduces to the standard expression for the net charge density $n = 4 \, \sinh(\xi)\, 
\n0$. On the other hand, one may consider two linear 
combinations of \rftwo{ncon}{wcon} leading to conservation equations 
of the form $\p_\mu \left[(n \pm w) \umU\right] = 0$. Using \rftwo 
{nden}{w}, we find $n \pm w = 4 \, \sinh[(\mu \pm \Omega)/T]\, \n0$, 
which indicates that thermodynamic quantities 
corresponding to charge and spin of the particles couple. 
In fact, $\Omega$ can be interpreted as a chemical
potential related with spin. Interestingly, 
from a thermodynamic point of view, a system of particles 
with spin $\onehalf$ can be seen 
as a two component mixture of scalar particles with chemical
potentials $\mu \pm \Omega$.

The resulting scheme, i.e., Eqs.~(\ref{Tmncon1}), (\ref{ncon}) and (\ref
{wcon}), can be regarded as a minimal extension of the standard 
perfect-fluid hydrodynamics of charged particles, where all dynamic 
equations follow from the conservation laws. We note that  (\ref
{Tmncon1}), (\ref{ncon}) and (\ref{wcon}) form a closed system of 
equations, which facilitates the study of spin dynamics. We may 
first solve these equations and subsequently use this solution as 
the dynamic background for the spin dynamics. Because of this 
property, we dub them the {\it  equations for hydrodynamic (spin) 
background}.

%\medskip
{\it 5. Spin dynamics:}  Our approach is based on the conservation of angular momentum
in the form $\p_\lambda J^{\lambda, \mu\nu}=0$, where $J^{\lambda, \mu\nu} = L^{\lambda, \mu\nu} 
+ S^{\lambda, \mu\nu}$ with $L^{\lambda, \mu\nu}=x^\mu T^{\nu\lambda}- x^\nu T^{\mu\lambda}$
and $S^{\lambda, \mu\nu}$ being the spin tensor. Since the energy-momentum tensor $\TmnU$
is symmetric (see \rf{Tmn}), the spin tensor $\slmnU$  satisfies the conservation law~\cite{Hehl:1976vr},
\bel{spincon1}
\p_\lambda \slmnU = 0.
\eel
For $\slmnU$ we use the form \cite{Becattini:2009wh}
\bel{st11}
\slmnU = \!\!\int\!\!\f{d^3p}{2 (2\pi)^3 E_p} \, p^\lambda \, {\tr} \left[(X^+\!-\!X^-) \SmunuU \right] 
=  \frac{w u^\lambda}{4 \zeta}  \omega^{\mu \nu}. \nn \\
\eel
We note that \rf{st11} differs from that derived in~\cite {Becattini:2013fla}.  We find that the additional 
terms given in Eq. (42) of \cite {Becattini:2013fla} are inconsistent with both the vortex solution discussed below and
the conservation law\rfn{spincon1}. Hence, we employ \rfn{st11}, which leads to a self-consistent framework.

Using \rf{wcon} and introducing the rescaled 
polarization tensor $\omnUbar = \omnU/(2\zeta)$, we obtain
\footnote{We stress again that (\ref{spincon1}) and (\ref{st12}) do not imply that the 
	spin dynamics described by these equations is trivial. Numerical solutions of 
	the hydrodynamic equations with spin presented here will be studied in a 
	future publication. 
}
\bel{st12}
u^\lambda \p_\lambda \, \omnUbar \equiv \f{d\omnUbar }{d\tau} = 0,
\eel
with the normalization condition $\omnLbar \, \omnUbar = 2$. The 
tensor $\omnLbar$ can be decomposed in a way analogous to  \rf
{omunuL},  with the two rescaled four-vectors $\kmLbar = \kmL/(2 
\zeta)$ and $\omLbar = \omL/(2 \zeta)$, satisfying the constraints
\bel{kbarobarort}
\hspace{-0.7cm} \kbar \cdot u = 0, ~~ \obar \cdot u = 0, ~~ \kbar \cdot \obar = 0, ~~ \kbar \cdot \kbar - \obar \cdot \obar = 1,
\eel
which leave only four independent components in $\kmLbar$ and 
$\omLbar$. This is expected, since the condition\rfn {conONE} 
removes one degree of freedom and another is eliminated by the rescaling 
with $\zeta$. The latter is anyway  determined by
the hydrodynamic background equations.

The last condition in\rfn{kbarobarort}  is fulfilled by 
employing the parameterization
\bel{mn}
\kmLbar = \mmL \sinh(\psi), \quad \omLbar = \nmL \cosh(\psi).
\eel
The four-vectors $\mmL$ and $\nmL$ are space-like and normalized to $-1$,
\bel{mnnorm}
\mmL m^\mu = -1, \quad \nmL \nmU = -1.
\eel
Using\rfn{mn} in\rfn{st12} we then find two coupled equations
\bel{spinequations}
&& \f{d\mmL}{d\tau} \sinh(\psi) + \mmL \cosh(\psi) \f{d\psi}{d\tau} + \mnL a^\nu  \sinh(\psi) \umL \nn \\
&& \hspace{2cm} + \epsLmnbg u^\nu a^\beta n^\gamma  \cosh(\psi) = 0, \nn \\
&& \f{d\nmL}{d\tau} \cosh(\psi) + \nmL \sinh(\psi) \f{d\psi}{d\tau} + \nnL a^\nu  \cosh(\psi) \umL \nn \\
&& \hspace{2cm} + \epsLmnab u^\nu a^\beta m^\alpha  \sinh(\psi) = 0,
\eel
where $a^\mu =  d\umU/d\tau$.

Equations\rfn{spinequations} should preserve the normalization 
conditions\rfn{mnnorm} as well as the ortoghonality constraints $m 
\cdot u = n \cdot u = m \cdot n = 0$. It is straightforward to convince 
oneself that these  
conditions are satisfied during the evolution of the system, provided
they are satisfied on the initial hypersurface and  the following 
equation is fulfilled by the variable $\psi$,
\bel{psi}
\f{d\psi}{d\tau} = \epsLmnbg \mmU u^\nu a^\beta n^\gamma.
\eel

%\medskip
{\it 6. Vortex solution:}  In order to demonstrate how our framework 
works in practice, we consider a rigid  rotation of the fluid 
around the $z$-axis. The hydrodynamic flow is defined by the 
four-vector $u^\mu$ with the components
\bel{uvortex}
\hspace{-0.5cm} u^0 = \gamma, \quad u^1 = - \, \gamma \, \Ot \, y, \quad u^2 = \gamma \, \Ot \, x, \quad u^3 = 0,
\eel
where $\Ot$ is a positive constant, $\gamma = 1/\sqrt{1 - \Ot^2 r^2}$, and $r$ 
denotes the distance from the center of the vortex in the transverse 
plane, $r^2 = x^2 + y^2$. Due to the limiting light speed, the flow profile
\rfn{uvortex} may be realized only within a cylinder with the radius $R < 1/\Ot$. 
The total time (convective)  derivative takes the form
\bel{cdvortex}
\f{d}{d\tau} = \umU \p_\mu = -\gamma \Ot \left(y \f{\p}{\p x} - x \f{\p }{\p y} \right).
\eel
Equation\rfn{cdvortex} can be used to find the fluid acceleration
\bel{a}
a^\mu = \f{d \umU}{d\tau} = - \gamma^2 \Ot^2 (0, x, y, 0).
\eel
As expected the spatial part of\rfn{a} points towards the centre of 
the vortex, as it describes the centripetal acceleration. 

It is easy to see that the equations for the hydrodynamic background 
are satisfied if $T$, $\mu$ and $\Omega$ are $r$ dependent and proportional to the local
Lorentz-$\gamma$ factor, namely
\vspace{-0.25cm}
\bel{TmuOm}
T = T_0 \gamma, \quad \mu = \mu_0 \gamma, \quad \Omega = \Omega_0 \gamma,
\eel
with $T_0$, $\mu_0$ and $\Omega_0$ being constants. One possibility 
is that the vortex represents an unpolarized fluid with $\omnL=0$ and 
thus, with $\Omega_0=0$. Another possibility is that the 
particles in the fluid are polarized and $\Omega_0 \neq 0$. In the 
latter case we expect  the polarization tensor to have the structure
\bel{omnvortex}
\omnL =
\left( \begin{array}{cccc}
0 & 0   & 0     & 0  \\
0 & 0   & \Ot/T_0 & 0  \\
0 & -\Ot/T_0 & 0    & 0  \\
0 &  0   &  0   & 0
\end{array} 
\right), 
\eel
where the parameter $T_0$ has been introduced to keep $\omnL$ dimensionless. 
This form, when used in \rfm{kmuomu}, yields 
$\kmL = \Ot^2 (\gamma/T_0) \, (0, x, y, 0)$ and $\omL = \Ot 
(\gamma/T_0) \, (0,0,0,1)$. As a consequence, we find $\zeta = 
\Ot/(2 T_0)$, which, for consistency with the hydrodynamic background 
equations, implies
\vspace{-0.25cm}
\bel{OmtOm0}
\Ot = 2 \, \Omega_0.
\eel
The factor 2 is a consequence of the fact that we are dealing with spin-$\half$ particles.
It follows that $\kmLbar = \gamma \Ot r \, (0, x/r, y/r, 
0)$ and $\omLbar = \gamma \, (0,0,0,1)$, leading to $m_\mu = 
(0,x/r,y/r,0)$, $n_\mu = (0,0,0,1)$, $\cosh(\psi) = \gamma$, and 
$\sinh(\psi) = \gamma \Ot r$. With all these quantities determined, 
it is rather straightforward to show that Eqs.\rfn
{spinequations} are fulfilled. We observe that $d\psi/d\tau = 0$, since 
the four-vectors $m^\mu$ and $a^\mu$ are parallel. We also note that 
the polarization tensor given by \rf{omnvortex} agrees with the thermal 
vorticity, namely
\bel{FBvortex}
\varpi_{\mu\nu} = -\f{1}{2} \left(\p_\mu \beta_\nu - \p_\nu \beta_\mu \right)
\eel
as emphasized in \cite{Becattini:2009wh,Becattini:2013fla}.

At this juncture, one may ask which vortex solution, polarized or 
unpolarized, is realized in Nature. Within the present framework the 
answer is that both of them can be realized and this depends on the 
boundary and initial conditions imposed on the hydrodynamic 
evolution. Spin relaxation effects, not included in the present
framework, drive the system to the state of maximum entropy. 

%\medskip 
{\it 7. Boost-invariant, polarized fluid:} Our system of equations allows also
for boost-invariant solutions describing polarized fluids. Assuming 
a one-dimensional, boost-invariant flow
 $u^\mu = (t/\tau,0,0,z/\tau)$, where $\tau=\sqrt{t^2-z^2}$
is the longitudinal proper time, we find that the hydrodynamic background
equations for $m \ll T$ are satisfied if $\mu/T$ and $\Omega/T$ are
constant, while $T$ is given by the Bjorken solution 
$T=T_0 (\tau_0/\tau)^{-1/3}$, with $T_0$ and $\tau_0$
being the initial temperature and proper time, respectively. 
One of the forms of the polarization tensor that satisfies
\rf{st12} in this case is  
\bel{BIomega}
\omnLbar =  \epsLmnbg u^\beta v^\gamma, 
\eel
where  $v^\gamma = (z/\tau,0,0,t/\tau)$ \cite{Florkowski:2008ag}.
Since $u^\lambda \p_\lambda u^\mu=0$ and $u^\lambda \p_\lambda v^\mu=0$, 
\rf{st12} holds also in the massive case where, however, the equations of
the hydrodynamic background must be solved numerically.  

%\medskip
{\it 8. Summary and discussion:} In this work we have introduced a 
hydrodynamic framework, which includes the evolution of the spin density 
in a consistent fashion. Equations that determine  the dynamics of the system follow 
solely from conservation laws. Thus, they can be regarded as a 
minimal extension of the well established perfect-fluid picture. 

Our framework can be used to determine the space-time dynamics of 
fluid variables, now including also the polarization tensor, from 
initial conditions defined on an initial space-like hypersurface.  
This property makes them useful for practical applications in 
studies of polarization evolution in high-energy nuclear collisions 
and also in other physics systems exhibiting fluid-like, collective 
dynamics connected with non-trivial polarization phenomena. In 
particular, the possibility to study the dynamics of systems in
local thermodynamic equilibrium 
represents an important advance compared to studies, where 
global equilibrium was assumed. 

Straightforward generalizations of our approach are possible to
multi-component fluids, to systems obeying Fermi-Dirac statistics and to systems with magnetic 
fields. It is also of central interest to extend our scheme by 
including dissipative effects. Of particular importance in the present context 
is clearly the relaxation of spin degrees of freedom, e.g., by means of a spin-orbit coupling, that
would drive the system towards equilibrium polarization~\footnote{See a closely related discussion of this issue, which follows Eq.~(23) in Ref.~\cite{Montenegro:2017lvf}.}, which in global equilibrium  is defined by $\omega_{\mu\nu}=\varpi_{\mu\nu}$
\cite{Becattini:2013fla}. The presence of a polarization 
tensor introduces a preferred directions in space, which suggests 
that concepts of anisotropic hydrodynamics may be useful for  
further developments of our formalism.

\medskip
%%%%%%%%%%%%%%%%%%%%%%%%%%%%%%%%%%
\begin{acknowledgments}
%%%%%%%%%%%%%%%%%%%%%%%%%%%%%%%%%% 

W.F. and A.J. thank Francesco Becattini for interesting discussions which 
motivated this work and the Mainz Institute for Theoretical Physics 
(MITP) for its hospitality and support during the workshop
``Relativistic Hydrodynamics: Theory and Modern Applications'',  
10--14 Oct. 2016.  W.F. also thanks Krzysztof 
Golec-Biernat for clarifying discussions concerning \rf{Mpmexp}. 
E.S. was supported by the Helmholtz Association grant No. VH-NG-823 at TU Darmstadt. 
This research was supported in part by the ExtreMe Matter Institute EMMI at the GSI 
Helmholtzzentrum f\"ur Schwerionenforschung, 
Darmstadt, Germany and by the Polish National Science Center Grant 
No. 2016/23/B/ST2/00717.

%%%%%%%%%%%%%%%%%%%%%%%%%%%%%%%%%%
\end{acknowledgments}
%%%%%%%%%%%%%%%%%%%%%%%%%%%%%%%%%%

%%%%%%%%%%%%%%%%%%%%%%%%%%%%%%%%%%%%%%%%%%%%%%%%%%%%%%%%%%%%%%%%
%\bibliography{spinref}


%%%%%%%%%%%%%%%%%%%%%%%%%%%%%%%%%%%%%%%%%%%%%%%%%%%%%%%%%%%%%%%%
%%%%%%%%%%%%%%%%%%%%%%%%%%%%%%%%%%%%%%%%%%%%%%%%%%%%%%%%%%%%%%%%
%%%%%%%%%%%%%%%%%%%%%%%%%%%%%%%%%%%%%%%%%%%%%%%%%%%%%%%%%%%%%%%%
\end{document}